\documentclass[12pt]{article}
\usepackage[centertags]{amsmath}
\usepackage{graphicx}

\begin{document}

\title{Chaos in the Honeycomb Optical Lattice Unit Cell}
\author{Maxwell Porter and  L. E. Reichl\\
Center for Complex Quantum Systems and Department of Physics\\
The University of Texas at Austin, Austin, Texas 78712\\}

\date{\today}

\maketitle\

\begin{abstract}

Natural and artificial honeycomb lattices are of great interest because the band structure of these lattices, if properly constructed, contains a Dirac point.  Such lattices occur naturally in the form of graphene and carbon nanotubes. They have been created in the lab in the form of semiconductor 2DEGs, optical lattices, and photonic crystals.  We show that, over a wide energy range, gases (of electrons, atoms, or photons) that propagate through these lattices are Lorentz gases and the corresponding classical dynamics is chaotic.  Thus, honeycomb lattices are also of interest for understanding eigenstate thermalization and the conductor-insulator transition due to dynamic Anderson localization.  

\end{abstract}

\section{Introduction}

Artificial honeycomb lattices  have  been realized in a variety of physical systems and  support propagation of a variety of waves. Honeycomb lattices composed of carbon atoms, in the form of graphene sheets  and carbon nanotubes, occur naturally and support electron matter wave propagation \cite{saito,katsnel}. Honeycomb lattices that have also been patterned in two dimensional semiconductor materials (2DEGs) can also support electron wave propagation,  \cite{gibert, goswami}.   Honeycomb lattices have  been realized in optical lattices  and support the propagation of atomic matter waves \cite{grynberg,wu1,wu2,tarruell}.  In addition,  photonic crystals have been constructed with honeycomb structure and support interesting effects in electromagnetic wave propagation \cite{haldane,sepk1,sepk2}. One reason for the great interest in these honeycomb lattices is  the unusual band structure, which if properly constructed, can support a Dirac point.  

As we shall show below, there is another reason why honeycomb lattices might be particularly interesting. 
Over a wide energy range, gases (composed of electrons, rubidium atoms, or photons) that traverse the honeycomb lattice can be considered to be a Lorentz gas, and the dynamics is classically chaotic.  This means that the honeycomb lattice is an ideal system for studying eigenstate thermalization \cite{srednicki, deutsch,rigol} or the effects of dynamic Anderson localization on the conduction properties of the lattice \cite{basko}.

In the sections below, we focus on the dynamics of a dilute gas of rubidium atoms in a honeycomb optical lattice because, as we shall show, this system provides an ideal system for studying the classical-quantum correspondence in lattice systems and, in particular, the effect of chaos on  wave propagation in lattices \cite{luna,boretz}.  In subsequent sections, we shall focus only on the unit cell of a honeycomb optical lattice, and  study the classical-quantum correspondence in the unit cell.

The honeycomb optical lattice can be formed by superposing traveling waves  whose electric fields are given by   ${\bf E}(x,y)={\sum_{j=1}^3}{\hat \epsilon}_je^{i({\bf k}_j{\cdot}{\bf r}+{\phi}_j)}$, where 
${\bf k}_3=\sqrt{3}k_L{\hat y}$, ${\bf k}_1=\frac{\sqrt{3}}{2}k_L{\hat y}-\frac{1}{2}k_L{\hat x}$, and ${\bf k}_2=\frac{\sqrt{3}}{2}k_L{\hat y}+\frac{1}{2}k_L{\hat x}$, $k_L$ is the wave vector of the radiation, ${\hat \epsilon}_j$ is the polarization of the $j$th wave,   and ${\phi}_j$ denotes the phases of the waves.
One wave travels along the y-axis and the remaining two waves propagate at $30^o$ angles on either side of the y-axis.  
The Hamiltonian describing the center of mass motion of Rubidium atoms in this radiation field   is given by
\begin{equation}
H=\frac{p_x^2}{2m_{Rb}}+\frac{p_y^2}{2m_{Rb}}+\frac{d^2|E(x,y)|^2}{{\hbar}{\Delta}},
\end{equation}
where $|E(x,y)|^2$ is the intensity of the radiation and  determines the potential energy experienced by the Rubidium atoms due to their interaction with the laser fields  \cite{graham, holder}.  It can be written
\begin{eqnarray}
|E(x,y)|^2=|E_0|^2 {\bigg[}3+2
  {\hat \epsilon}_1{\cdot}{\hat \epsilon}_2 \cos (k_L x+{\phi}_{21}) ~~~~~~~~~~~~~~~~~~~~~~~~~~~~~~~~~~~ \nonumber\\
  +2 {\hat \epsilon}_1{\cdot}{\hat \epsilon}_3 \cos \left(\frac{ \sqrt{3}}{2} k_L y+ \frac{k_L x}{2}+{\phi}_{31})\right)
+2 {\hat \epsilon}_2{\cdot}{\hat \epsilon}_3\cos
   \left(\frac{ \sqrt{3}}{2} k_L y-\frac{k_L x}{2}+{\phi}_{32}\right){\bigg]}~~~~~~~~~
 \label{pot1}
\end{eqnarray}
where ${\phi}_{ij}={\phi}_i-{\phi}_j$ denote the relative phases of the waves.
The depth of the potential well $V(x,y)$   is proportional to the laser intensity, and is determined by the angle between the polarization vectors of the radiation fields.  Each  polarization vector ${\hat \epsilon}_j$ is  perpendicular to the direction of propagation ${\bf k}_j$ of its radiation field, but  ${\hat \epsilon}_j$ can be rotated about that axis. 
The polarization vectors can be written ${\hat \epsilon}_j={\cos}({\phi}_j){\sin}({\theta}_j){\hat x}+{\sin}({\phi}_j){\sin}({\theta}_j){\hat y}+{\cos}({\theta}_j){\hat z}$, with ${\phi}_1=120^o$,  ${\phi}_2=60^o$, and  ${\phi}_3=0^o$.  Then the coefficients in Eq. (\ref{pot1}) can be written
${\alpha}_1={\epsilon}_1{\cdot}{\epsilon}_2={\cos}({\theta}_1){\cos}({\theta}_2)+\frac{1}{2}{\sin}({\theta}_1){\sin}({\theta}_2)$,
${\alpha}_2={\epsilon}_1{\cdot}{\epsilon}_3={\cos}({\theta}_1){\cos}({\theta}_3)-\frac{1}{2}{\sin}({\theta}_1){\sin}({\theta}_3)$, and 
${\alpha}_3={\epsilon}_2{\cdot}{\epsilon}_3={\cos}({\theta}_2){\cos}({\theta}_3)+\frac{1}{2}{\sin}({\theta}_2){\sin}({\theta}_3)$. 

Let us now introduce dimensionless variables, $x'$, $y'$, $t'$, $H'$, $U'$, $E'$,  where $x'=k_Lx$, $y'=k_Ly$, $t'={\omega}_Lt$, $H=H'E_L$, $U=\frac{d^2|E_o|^2}{{\hbar}{\Delta}}=U'E_L$, $p_x={\hbar}k_Lp_x'$, and  $p_y={\hbar}k_Lp_y'$. If we rewrite the Hamiltonian in terms $x'$, $y'$, $t'$, $H'$, $E'$ and then drop the primes, the Hamiltonian can be written
\begin{equation}
H=p_x^2+p_y^2+UV(x,y) =E-\frac{3}{2}U,
\label{ham1}
\end{equation}
where,  
\begin{equation}
V(x,y)={\alpha}_1  \cos (x)+{\alpha}_2 \cos \left(\frac{ \sqrt{3}y}{2} + \frac{ x}{2}-{\pi}\right)+{\alpha}_3 \cos
   \left(\frac{ \sqrt{3}y}{2}-\frac{x}{2}-{\pi}\right)
\label{ham2}
\end{equation}
and we have made a particular choice of the phases ${\phi}_{21}=0,~{\phi}_{31}={\phi}_{32}=-{\pi}$ to give a convenient orientation of the lattice unit cell with respect to the coordinate frame.  
This Hamiltonian neglects the interactions between the rubidium atoms, which form a dilute gas, and only accounts for the atom-radiation interaction.   For gas  comprised of rubidium atoms,  $m_{Rb}=86.909~{\rm u}$(the mass of $^{87}Rb$).   The recoil energy of rubidium is $E_L={\hbar}{\omega}_L=\frac{{\hbar}^2k_L^2}{2m_{Rb}}=2.156{\times}10^{-30}~{\rm J}$ so ${\omega}_L=2.044{\times}10^4~{\rm rad/s}$.

The Hamiltonian in Eqs. (\ref{ham1}) and (\ref{ham2}) has an important scale invariance.    
Let us change the intensity of the laser radiation so that the new intensity ${\tilde U}$ is proportional  to the old intensity $U$ with ${\tilde U}={\beta}U$.  For the case ${\beta}>1$ (${\beta}<1$)  this will increase (decrease) the height of the optical lattice potential energy and change the energy scale of the dynamics.  However, let us now make the following changes in the variables: $E={\beta}{\tilde E}$, $H={\beta}{\tilde H}$, $p_x=\sqrt{\beta}{\tilde p}_x$, $p_y=\sqrt{\beta}{\tilde p}_y$, and $t={\tilde t}/\sqrt{\beta}$. Then the Hamiltonian in Eq. (\ref{ham1}) takes the form
\begin{equation}
{\tilde H}={\tilde p}_x^2+{\tilde p}_y^2+UV(x,y) ={\tilde E}-\frac{3}{2}U,
\label{ham3}
\end{equation}
Also, Hamilton's equations remain the same, except they are expressed in terms of the variables $\{{\tilde p}_x,{\tilde p}_y, x,y,{\tilde t},U \}$ rather than $\{{p}_x,{p}_y, x,y,{ t},U \}$.  This scaling properly is very important for the quantum-classical correspondence of the system because action variables associated with periodic orbits scale as $\sqrt{\beta}$.  Since semi-classical quantization stipulates that action is quantized in units of Planck's constant $h$ \cite{born}, it means that the number of quantum states that a given periodic orbit in the classical phase space can support, increases as  $\sqrt{\beta}$.  As we scale the laser intensity, the spatial scale and identity of dynamical structures, such as periodic orbits and KAM islands don't change, but their energy and the number of quantum states the support does change.  Thus, this is a perfect system for studying the quantum-classical correspondence for wave motion through a periodic lattice.    In subsequent sections, we will analyze the dynamics for   $U=20$, a value attainable in current experiments \cite{greiner02}, and a value for which localized states can exist in the graphene optical  lattice.  

When the polarizations of all three waves are parallel, for example ${\theta}_1={\theta}_2={\theta}_3=0$ so  ${\alpha}_1= {\alpha}_2={\alpha}_3=1$,   the optical lattice has a perfect hexagonal structure similar to graphene.    In Fig. 1.a, we show a contour plot  of the potential energy $V(x,y)$ for the case  ${\alpha}_1= {\alpha}_2={\alpha}_3=1$.  We have indicated the lattice unit cell of  graphene,  by the dashed lines.   The  primitive vectors for the graphene unit cell are  $\mathbf{a}_{1}=2{\pi}{\hat {e}_x}+\frac{2{\pi}}{\sqrt{3}}{\hat {e}_y}$ and $\mathbf{a}_{2}=2{\pi}{\hat {e}_x}-\frac{2{\pi}}{\sqrt{3}}{\hat {e}_y}$ and the lattice constant is $|\mathbf{a}_{1}|=|\mathbf{a}_{2}|=\frac{4{\pi}}{\sqrt{3}}$. The unit cell  has an area of $\Omega =\frac{8{\pi}^2}{\sqrt{3}}$.

It is interesting to consider what happens to the lattice when the laser polarizations are not perfectly aligned.  In Fig. 1.b, we show a contour plot of the lattice potential energy for the case ${\theta}_1=0$, ${\theta}_2=30^o$ and ${\theta}_3=60^o$, so that ${\alpha}_1=\frac{\sqrt{3}}{2}$,  ${\alpha}_2=\frac{1}{2}$, and  ${\alpha}_3=\frac{3\sqrt{3}}{8}$. All the maxima and  the five saddle points   of the unit cell and  remain in the same position.   However, the interior of the unit cell becomes twisted relative to that of graphene and the position of the two  potential energy minima inside the unit cell changes.

\section{Classical Dynamics}

The classical dynamics of the optical lattice gives important insight into the behavior of the quantum system.  The ``unit cell"  for the classical dynamics is one-half of the graphene unit cell described above - for example the equilateral triangle on the right half of the graphene unit cell - which we shall call the {\it  half-cell}. The structure of the classical dynamics repeats itself in each of these  half-cells.
In subsequent sections we will focus the classical dynamics in the right half-cell.  The classical  half-cell forms a symmetric equilateral triangle and has potential energy maxima at each corner of the triangle. It has a saddle at the center of each boundary line, and it has a potential energy  minimum at its center.  

We can determine the location of the dominant  fixed points of the lattice from 
Hamilton's equations, which can be written
\begin{eqnarray}
\frac{dp_x}{dt}=U{\alpha}_1\sin (x)-\frac{1}{2} U{\alpha}_2 \sin \left(\frac{x}{2}+\frac{\sqrt{3}
   y}{2}\right)-\frac{1}{2}U{\alpha}_3\sin \left(\frac{x}{2}-\frac{\sqrt{3}
   y}{2}\right),~~~\nonumber\\
\frac{dp_y}{dt}=-\frac{ \sqrt{3}}{2} U{\alpha}_2 \sin
   \left(\frac{x}{2}+\frac{\sqrt{3} y}{2}\right)+\frac{ \sqrt{3}}{2} U{\alpha}_3 \sin \left(\frac{x}{2}-\frac{\sqrt{3}
   y}{2}\right),~~~\nonumber\\
\frac{dx}{dt}=2p_x,~~~   \frac{dy}{dt}=2p_y.~~~~~~~~~~~~~~~~~~~~~~~~~
\end{eqnarray}
 Fixed points are points for which ${\dot p}_x={\dot p}_y={\dot x}={\dot y}=0$.  There are several  fixed points  of these equations that  are independent of the values of $U$ and ${\alpha}_j$, as can be seen in Figs. 1.a and 1.b.   These include the potential energy maxima  and the saddle points.  The potential energy maxima occur at  the corners of the half-cell $(x_{mx}=2{\pi},~y_{mx}=0)$ and  $(x_{mx}=0,~y_{mx}={\pm}\frac{2{\pi}}{\sqrt{3}})$, and have energy $E_{mx}=({\alpha}_1+{\alpha}_2+{\alpha}_3)U+\frac{3U}{2}$. There are three saddle points associated to the half-cell.   The saddle point  located at $(x_{sd1}=0,~y_{sd1}=0)$ has energy $E_{sd1}=({\alpha}_1-{\alpha}_2-{\alpha}_3)U+\frac{3U}{2}$. The saddle points  located at {$(x_{sd{\pm}}={\pi},~y_{sd{\pm}}={\pm}\frac{{\pi}}{\sqrt{3}})$ have energy $E_{sd{\pm}}=(-{\alpha}_1{\pm}{\alpha}_2{\mp}{\alpha}_3)U+\frac{3U}{2}$.  For graphene (${\alpha}_1={\alpha}_2={\alpha}_3=1$), the saddle points have equal energy $E_{sd1}=E_{sd{\pm}}=10$ .  For other values of ${\alpha}_i$, the saddle points, which control the flow of trajectories through the lattice, have different  energy.  
 
 There is one  potential energy minimum  inside  the half-cell.
 For the case of graphene (${\alpha}_1={\alpha}_2={\alpha}_3=1$), the fixed point at the potential energy minimum,  is  located at  $(x_{mn}=\frac{2{\pi}}{3},~y_{mn}=0)$, and it has energy $E_{mn}=0$.  For the lattice shown in Fig. 1.b, the fixed point at the potential energy minimum is located at $(x_{mn}=2.53,~y_{mn}=0.45)$ and the potential energy is $E_{mn}=-1.08$.

  For the graphene-like optical lattice, the classical dynamics for the energy interval below the saddle point energy  $(0{\leq}E{\leq}10)$ has a mixed phase space. We can visualize the dynamics with Poincare surfaces of section \cite{reichl}.  Each half-cell has three straight lines that are local minima of the potential energy (minimum potential energy ``trenches").  They start at the saddle points and end at the potential energy minimum at the center of the half-cell.  We can use Birkhoff coordinates to obtain Poincare surfaces of section (SOSs) along each of these trenches, and they will be identical, due to the symmetry of the lattice.   All SOSs we show here are plots of $p_x$ versus $x$, plotted each time the trajectory crosses   the line $y=0$ with positive $p_y$.  While only the portion of this line $0<x<2\pi/3$ is a trench, we can extend the line $y=0$  across the full width of the half cell to a more SOS. 
   In Fig. 2. we show the progression of the dynamics for surfaces of section along the $y=0$ trench in the unit half-cell.  In Fig. 2.a, we show the SOS for energy $E=5.338$.  It is dominated by KAM tori.  All SOSs in  an approximate energy range $0<E{\leq}5.5$   have this same large scale  structure and simply grow in size as the energy increases. Above energy $E{\approx}5.5$   the chaos begins to spread, as can be seen in Fig. 2.b, and as we reach the saddle point energy the phase space is dominated by chaos.       

Figures 2.a and 2.b show the dominant stable and unstable periodic orbits of the SOS.  The corresponding configuration space orbits are shown in Fig. 3.  There are two dominant period-one stable periodic orbits in Fig. 2.a, located at $(\frac{p_x}{\sqrt{E}}=0,x=1.55)$ and $(\frac{p_x}{\sqrt{E}}=0,x=2.74)$.  These are due to the configuration space periodic orbit in Fig. 3.a and its time-reversed twin.  This periodic orbit, and its time reversed twin, each  undergo a bifurcation at energy $E=9.09$  and  give rise to the pairs of small stable islands shown in the middle and on the right side of Fig. 2.c.  The configuration space version of the bifurcated orbit is shown in Fig. 3.b.  There are three unstable period-one periodic orbits in Fig. 2.a, located at  $(\frac{p_x}{\sqrt{E}}={\pm}0.66,x=1.60)$, and $(\frac{p_x}{\sqrt{E}}=0,x=2.38)$ coming from the three independent unstable periodic orbits in Fig. 3.c.  
If one moves slightly off of these unstable periodic orbits, the trajectory lies in the chaotic sea and  undergoes a random walk through the phase space, as shown in Fig. 3.d.  This random walk occurs mostly in the neighborhood of the periodic orbits, but the trajectory eventually visits the neighborhood of all three periodic orbits.

When the phase space is predominantly integrable, and large regions of KAM tori exist, we can hope to use semi-classical quantization to determine the existence of quantum states in this region.  The semi-classical quantization condition allows us to determine if the potential well below the saddle can hold a quantum state. To hold a quantum state, a classical orbit must satisfy the action relationship $J=nh$, where J is the action (equal to the area enclosed by the orbit in phase space), $h$ is Planck's constant, and n is an integer. The dominant period-one stable periodic orbits satisfy this condition at energy  $E\approx5.473$.  This periodic orbit, and its time reversed twin, appear to allow the creation of a standing wave bound state of the lattice (this energy differs only  slightly from the  symmetric quantum ground state shown in Fig. 7 and with $E\approx5.30$).   Above these energies, the growing regions of chaos, and destruction of action as a good quantum number, preclude the use of semi-classical quantization \cite{reichl}. 
 
For energies above the saddle point energy, there is chaotic flow throughout the lattice.  In the  half-cell, trajectories can enter and leave the half-cell through all three saddle points.  The mixing of the chaotic flow from the three saddle points can be seen in  Fig. 4. The saddle-points themselves have energy  $E=10$.  However, the  saddle provides an unstable period-one orbit in the SOS for energies $10{\leq}E<90$, where $E=90$ is the energy of the potential peak.  In Fig. 4.a,  we show the unstable manifold coming off the saddle at $(x=0,y=0)$ and energy $E=12.5$.  The saddles have a stable direction, along which the orbits oscillate periodically, and an unstable direction in which the orbits (roll down the hill) move away from the saddle point with growing speed.  An unstable manifold at energy $E=12.5$ has an oscillatory motion as it moves away from the fixed point  that allows for a clean SOS.  The unstable manifold is area preserving and crosses and recrosses its own saddle region as it evolves. It appears to fill the chaotic region of the phase space (see Fig. 4.c).  In Fig. 4.b we show the same surface of section, but now include the flow of the stable manifolds that originate from the saddles at $(x=\pi,{\pm}\frac{\pi}{\sqrt{3}})$.  The unstable manifolds from all three saddle points mix together and flow back and forth across the saddle at $(x=0,y=0)$.  If we did SOSs along the minimum potential energy trenches that start from the saddle points at   $(x=\pi,{\pm}\frac{\pi}{\sqrt{3}})$, we would see identical behavior. This gives a clear indication of the highly mixing behavior of the chaotic flow through the optical lattice.  Indeed, trajectories in the chaotic sea undergo a random walk through the optical lattice, even though the Newtonian dynamics is completely deterministic.

For energies above the saddle point energy, $10{\leq}E{\leq} 90$,  the atoms moving though the optical lattice ``see" potential barriers arranged like that of  a Lorenz gas, whose dynamics is known rigorously a K-flow \cite{geisel, gaspard1}.  The unit half-cell, itself,  is very similar to the three disk system, which is a classic example of a chaotic system.  When the three disks stand alone, the three disks  form  a chaotic scattering system (K-flow)  \cite{eckhardt,gaspard2}.  Scattering for three ``soft" disks, with circular  potential barriers somewhat similar to our optical lattice, has been studied by Jung and Richter \cite{jung}.  They find an energy interval  for their soft potential for which the scattering {\it is} a K-flow.   For the energy interval $E_{sd}=10$ to about $E{\approx}32$ in the honeycomb optical lattice, the SOS is dominated by chaos, although small stable islands do appear in the SOS.  For the energy range  $32<E<55$, no stable islands appear in the SOS, as can be seen from the SOS in Fig. 5.a, and the dynamics may well  be a true K-flow, in analogy to the behavior found in \cite{jung}. Above  $E \approx 55$, three stable fixed points  emerge in the SOSs.  At slightly higher energies,  three additional, but much smaller, stable fixed points appear in the SOS. These  six stable fixed points begin to dominate the SOS  at energy $E_{peak}=90$ as can be seen in Fig. 5.b.  They are the result of periodic orbits that exist in the configuration space at these higher energies.  The two  configuration space periodic orbits that give rise to the six dominant stable fixed points in Fig. 5.b,  are shown in Fig. 3.e.  As we go higher in energy, larger regions of chaos appear and then gradually disappear at very high energies. Indeed, there still is significant chaos at $E=360$.  This complex structure of the phase space, for energies above $E=90$, is the result of higher order resonances that exist at these high energies.

\section{Quantum Dynamics}

Each unit cell of graphene consists of two half-cells of the classical system.  
If we consider only the dynamics of the unit cell, the energy eigenstates can be written 
\begin{equation}
 u_E({\bf r})=\frac{\sqrt{3}}{8{\pi}^2}{\sum_{n_1,n_2=-\infty}^{\infty}}A_{E;n_1,n_2}{\rm e}^{i(n_1{\bf b}_1+n_2{\bf b}_2){\cdot}{\bf r}}
\end{equation}
where  $\mathbf{b}_{1}=\frac{1}{2}{\hat {e}_x}+\frac{\sqrt{3}}{2}{\hat {e}_y}$ and
$\mathbf{b}_{2}=\frac{1}{2}{\hat {e}_x}-\frac{\sqrt{3}}{2}{\hat {e}_y}$ are the reciprocal lattice vectors.
The eigenvalue equation in a unit cell then takes the form
\begin{equation}
{\hat H}u_{E}({\bf r})=\left(\frac{\hbar ^{2}}{2m} \mathbf{\nabla } ^{2}+V\left( \mathbf{r}\right)\right) u_{E}({\bf r}) =Eu_{E}({\bf r}) %
\end{equation}
The boundary condition for $u_{\mathbf{k}}\left( \mathbf{r}\right) $ is $u_{%
\mathbf{k}}\left( \mathbf{r+a}_{i}\right) =u_{\mathbf{k}}\left( \mathbf{r}%
\right) $, $i=1,2$.
We can  write $u_{\mathbf{k}}\left( \mathbf{r}\right) $ as
\begin{equation}
u_{\mathbf{k}}\left( \mathbf{r}\right) =\frac{1}{\sqrt{\Omega }}\sum_{\mathbf{G}}A_{\mathbf{G}}e^{i\mathbf{G\cdot r}} 
 \label{uk in plane wave}
\end{equation}
where $\mathbf{G}=n_1\mathbf{b}_{1}+n_2\mathbf{b}_{2}$ ($n_1,\;n_2=0,\;\pm 1,\;\pm
2,\;\cdots ,\;\pm \infty $), and $A_{\mathbf{G}}$ are coefficients to be determined. 
The basis set has been normalized so that
\begin{equation}
{\frac{1}{{\Omega }}}\int_{\Omega }d\mathbf{r}e^{-i(\mathbf{G}^{\prime }-%
\mathbf{G})\mathbf{\cdot r}}=\delta _{\mathbf{G}^{\prime }\mathbf{,G}}.
\label{orthonormal}
\end{equation}
After we substitute Eq. (9) into Eq. (8)
then,  with the aid of Eq. (\ref{orthonormal}), we can write
the Schrodinger equation in matrix form:
\begin{equation}
\sum_{\mathbf{G}^{\prime }}\left[ \frac{\hbar ^{2}}{2m}G^{2}\delta _{\mathbf{G,G}^{\prime }}+V_{\mathbf{G,G}%
^{\prime }}\right] A_{\mathbf{G}^{\prime }}=E_{\mathbf{k}}A_{\mathbf{G}},
\label{mateig1}
\end{equation}

In Fig. 6,  we show the wavefunctions (to within an overall phase factor) of the  two lowest energy eigenstates (with energies $E=5.30$ and $E=5.31$) of the  unit cell. The ground state of the lattice ($E=5.30$) is symmetric in the unit cell. The first excited state ($E=5.31$ is antisymmetric in the unit cell.  These two lowest states are  the only eigenstates with energy below the saddle point energies, and may be thought to correspond to the $\pi$-bonds of graphene. The next higher energy eigenstates have energies just above the saddle energies, and  form two degenerate pairs of states.   The wave functions for the degenerate states with energy $E=10.09$ are shown in Figs. 7.a and 7.b. This pair of degenerate states is clearly associated with the saddle points.   The wave functions for the degenerate states with energy $E=10.25$ are shown in  Figs. 7.c and 7.d. These states appear to be excited states of the potential wells in the unit cell.

In Fig. 8.a, we show an energy eigenstate for energy $E=57.79$ where the phase space is in the chaotic regime.  The unit cell is indicated by the dashed lines.  For the laser intensity we are considering, $U=20$, the system is far from the semi-classical regime. However, some of the signatures of chaos can be seen.    The state is symmetric about the line $x=0$ and it is anti-symmetric about the line $y=0$.  In Fig. 8.b, we focus on the half-cell.  There is a six-fold symmetry for this state, but within each of the six triangles in the half-cell, we begin to see the irregular nodal patterns characteristic of eigenstates in a chaotic system.  As we increase the laser intensity, the energy of these states will scale upward in the manner described earlier, and we expect to see a denser irregular pattern of nodal lines, similar to chaotic billiards like the stadium or the Sinai billiard \cite{reichl}.

\section{Conclusions}

We have analyzed the classical and quantum dynamics of the unit cell of a honeycomb optical lattice.
The honeycomb optical lattice is of particular interest because, with the proper scaling, the dynamical structure of the phase space remains unchanged as the energy of the system is changed.  This means that it is possible to go from the quantum regime to the semiclassical regime, without changing the basic dynamics of the system. Below the saddle, classical trajectories are localized.  Above the saddle, the system consists of an array of circular barriers, and the dynamics of particles (electrons, atoms, photons) confined to the lattice is that of a Lorentz gas.  
The particle dynamics over a wide energy range is chaotic.  

For a laser intensity commonly found in  optical lattice experiments, we  have found that the unit cell  of the honeycomb lattice contains a symmetric and antisymmetric pair of ground states  below the low energy saddle.  These ground states are well separated in energy from all other energy eigenstates, all of which lie above the saddle.  

This system, because it can be scaled from the quantum regime to the semi-classical regime without changing the qualitative structure of the dynamics, provides an important system for studying the effects of lattice dynamics on the Dirac point, the validity of the ``eigenstate thermalization hypothesis", and  the possible occurrence of dynamic Anderson localization as a mechanism for the conductor-insulator transition in the lattice.

 \vspace{0.2cm}

{\bf Acknowledgements}
The authors thank the Robert A. Welch Foundation (Grant No. F-1051) for support of this work.  We also thank Christof Jung for helpful comments.

\newpage

\subsection{List of Figures}

%fig.1
\begin{figure}[!hp]
\centering
 \scalebox{.9}{\includegraphics{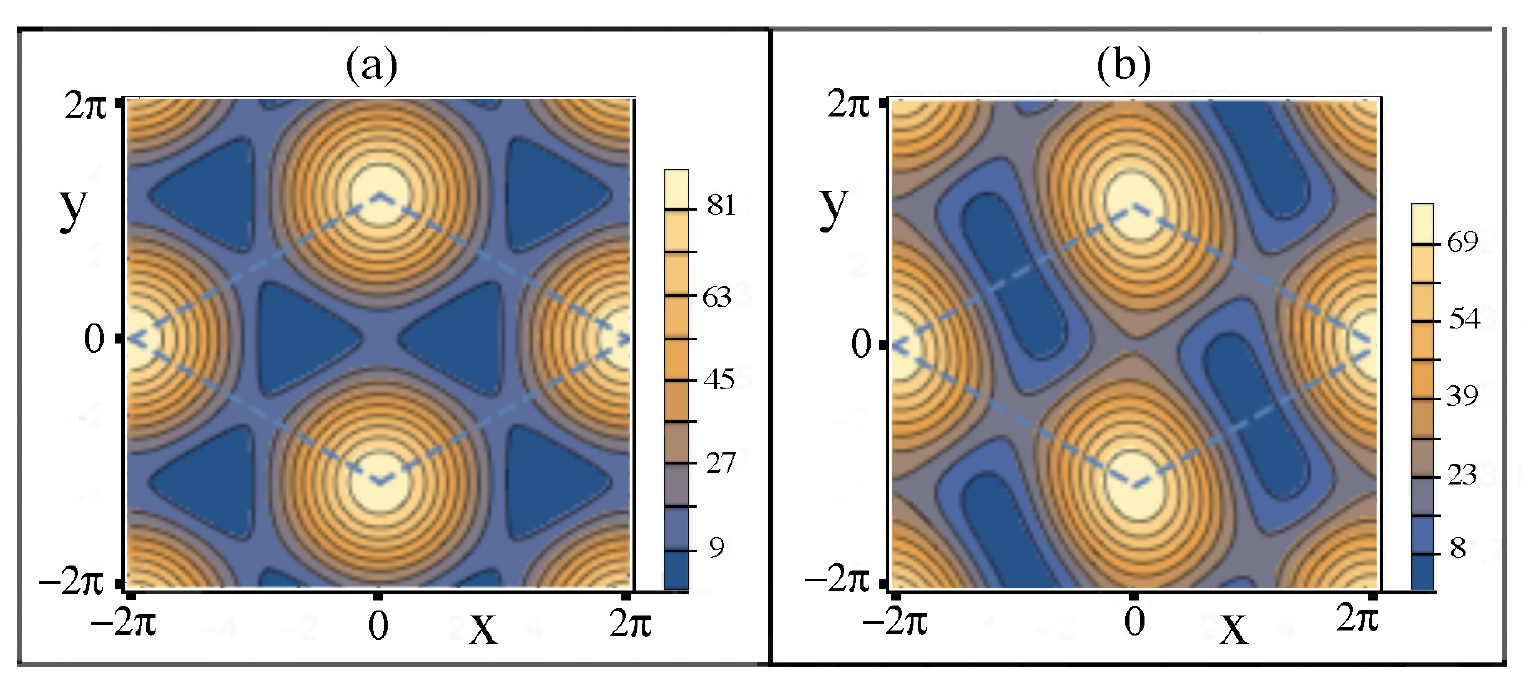}}
\caption{(a) Contour plot of the potential energy for $U=20$ and ${\theta}_1= {\theta}_2={\theta}_3=0^o$ (graphene-like honeycomb).
Contour plot of the potential energy for $U=20$ and ${\theta}_1=0^o$, $ {\theta}_2=30^o$, and ${\theta}_3=60^o$ (twisted honeycomb).  }
\label{fig:poten}
\end{figure}
%

%fig.2
\begin{figure}[!hp]
\centering
 \scalebox{.8}{\includegraphics{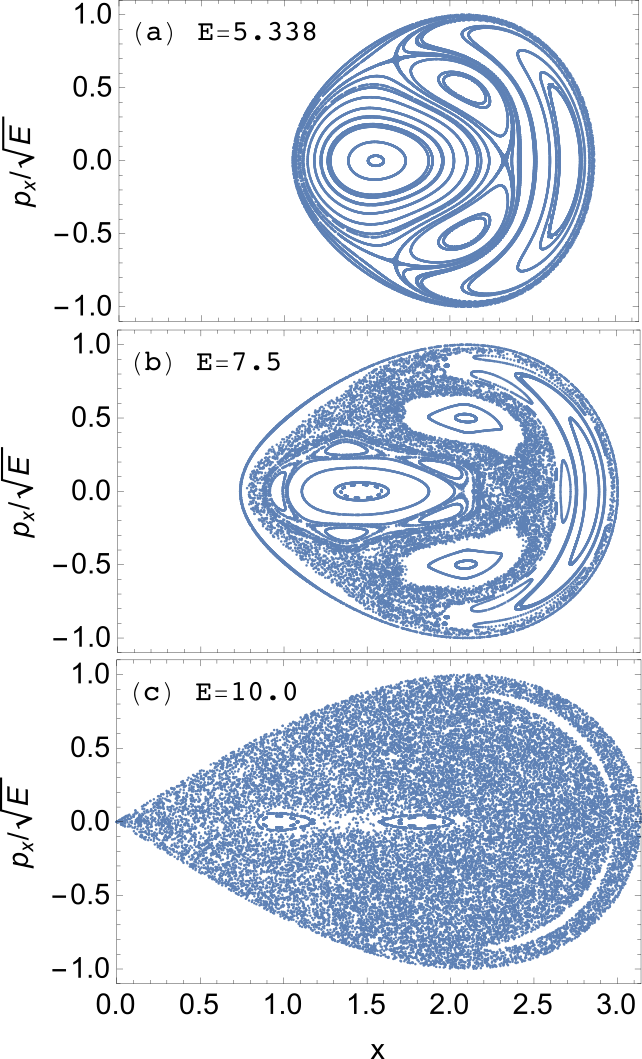}}
\caption{Surfaces of Section of $(p_x,x)$ for $y=0$ and $p_y>0$ below and at the saddle  energy. (a) $E=5.338$. (b) $E=7.5$. (c) E=$10.0$.  }
\label{fig:lowsos}
\end{figure}
%

%fig.3
\begin{figure}[!hp]
\centering
 \scalebox{.6}{\includegraphics{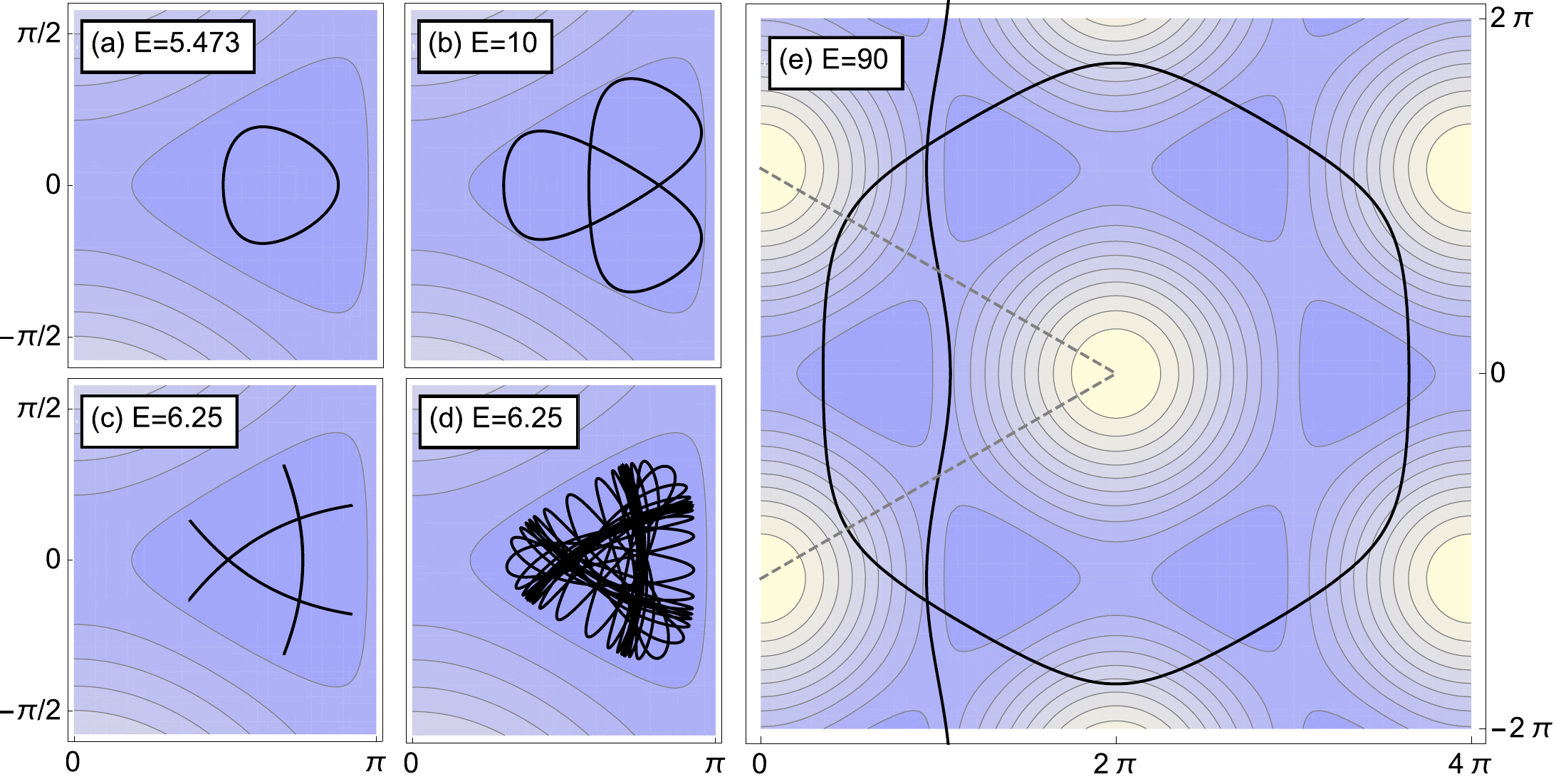}}
\caption{Configuration space plots of key periodic orbits. (a) Dominant periodic orbit below the saddle point energy, $E=5.473$. (b) Bifurcated orbit in (a) but at energy $E=10$.  (c) Three unstable periodic orbits at $E=6.25$. (d) Trajectory in chaotic sea at $E=6.25$. (e) Two dominant periodic orbits at high energy, $E=90$.  }
\label{fig:porb}
\end{figure}
%

%fig.4
\begin{figure}[!hp]
\centering
 \scalebox{1}{\includegraphics{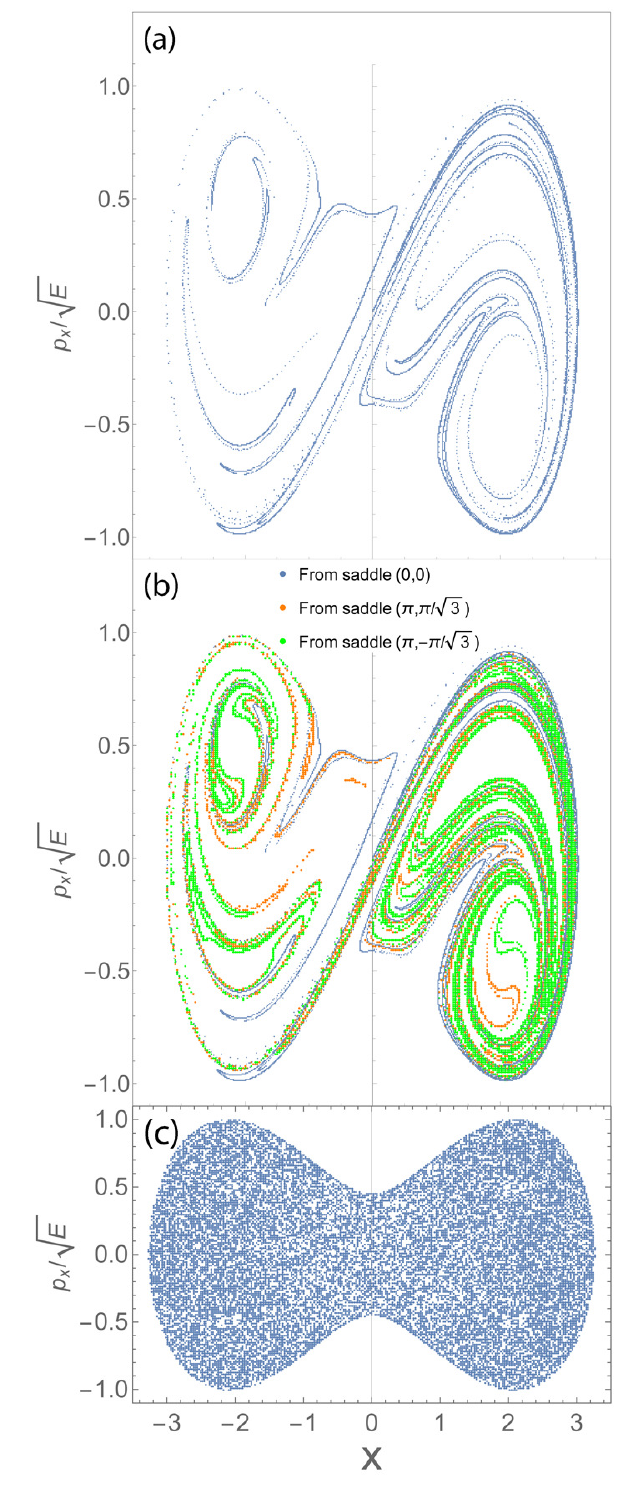}}
\caption{(a) SOS of $(p_x,x)$ for $y=0$ and $p_y>0$, of the unstable manifold that emerges from the saddle at $(x=0,y=0)$ for energy $E=12.5$. (b) SOS  of $(p_x,x)$ for $y=0$ and $p_y>0$ containing the unstable manifolds coming from all three saddles in the half-cell for $E=12.5$. (c)  Phase space SOS of  $(p_x,x)$ for $y=0$ and $p_y>0$ for a range of initial conditions in the half-cell for energy $E=12.5$. }
\label{fig:tang}
\end{figure}
%

%fig.5
\begin{figure}[!hp]
\centering
 \scalebox{1}{\includegraphics{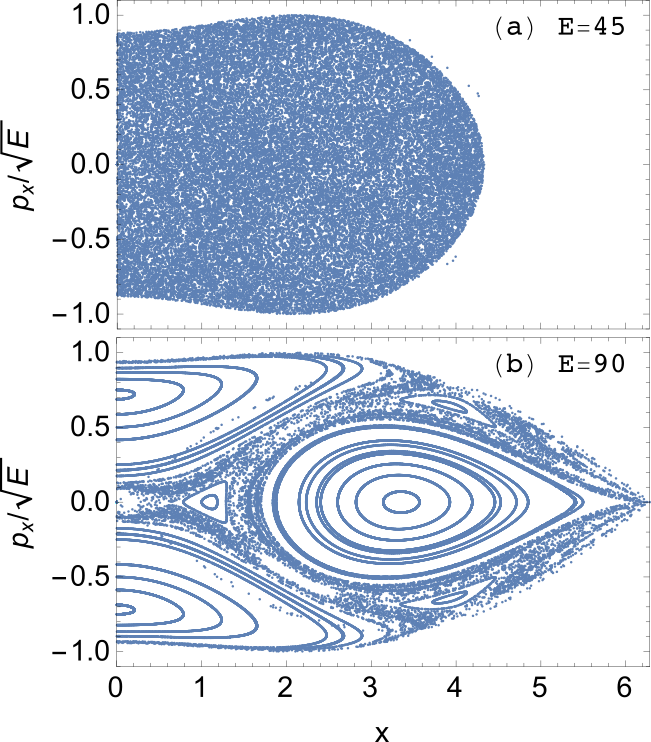}}
\caption{SOS of $(p_x,x)$ for $y=0$ and $p_y>0$ for (a) $E=45$ and (b) $E=90$. }
\label{fig:ucell}
\end{figure}
%

%fig.6
\begin{figure}[!hp]
\centering
 \scalebox{.8}{\includegraphics{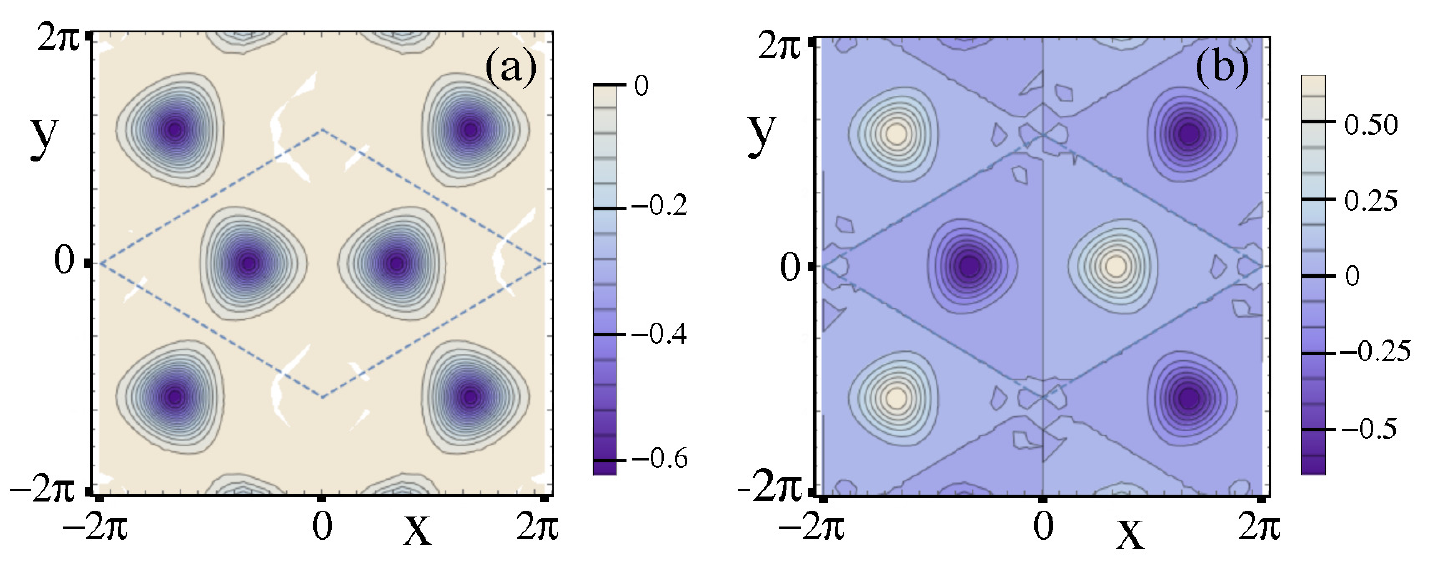}}
\caption{The wave functions for the lowest energy eigenstates in the unit cell. (a) E=5.30$. (b) $E=5.31  }
\label{fig:quant1-2}
\end{figure}
%

%fig.7
\begin{figure}[!hp]
\centering
 \scalebox{.9}{\includegraphics{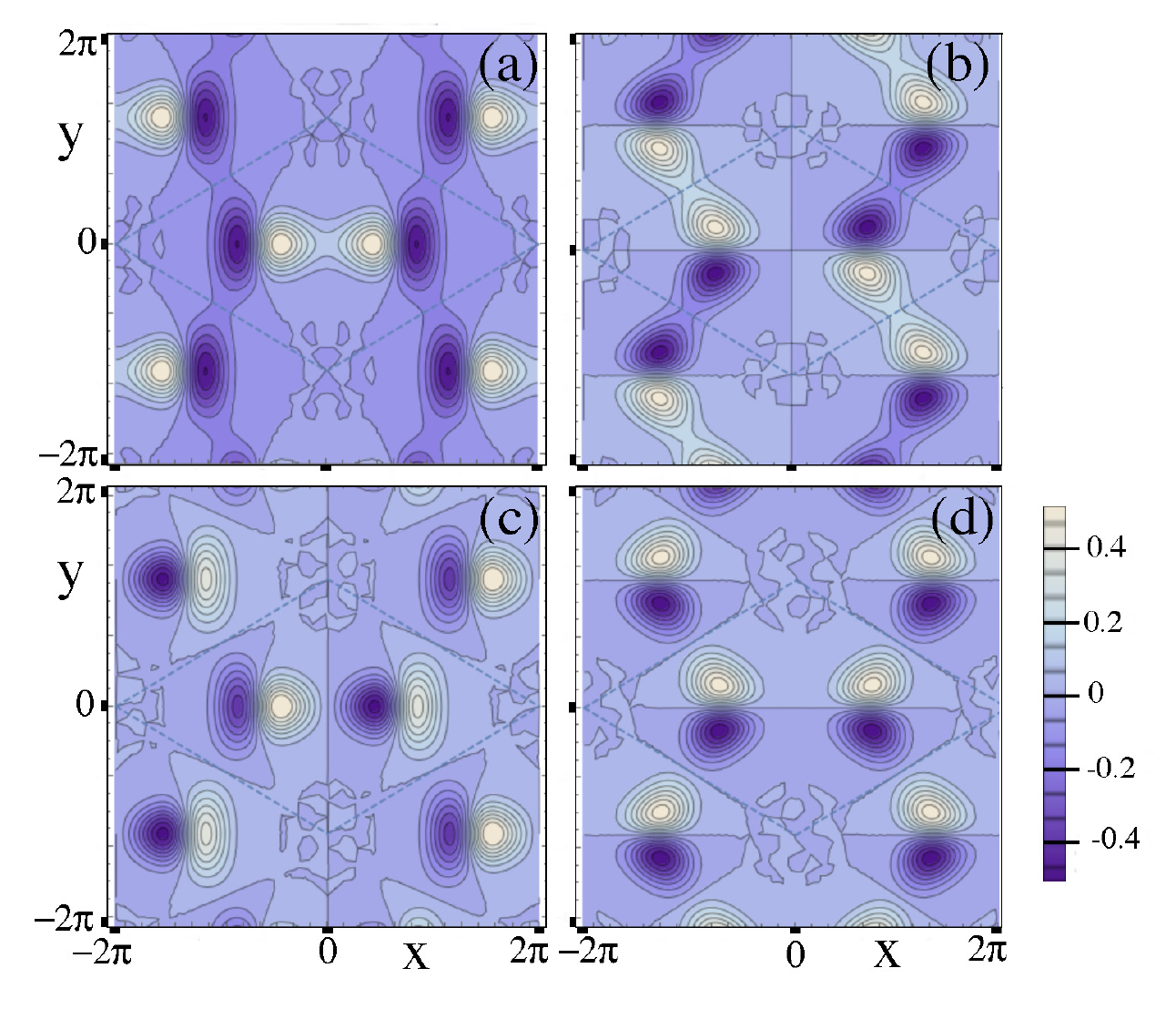}}
\caption{The wave functions for the  third through sixth energy eigenstates.  They form degenerate pairs.  (a) and (b) have energy $E=10.09$ which is just above the saddle. (c) and (d) have energy $E=10.25$.  }
\label{fig:quant3-6}
\end{figure}
%

%fig.8
\begin{figure}[!hp]
\centering
 \scalebox{.9}{\includegraphics{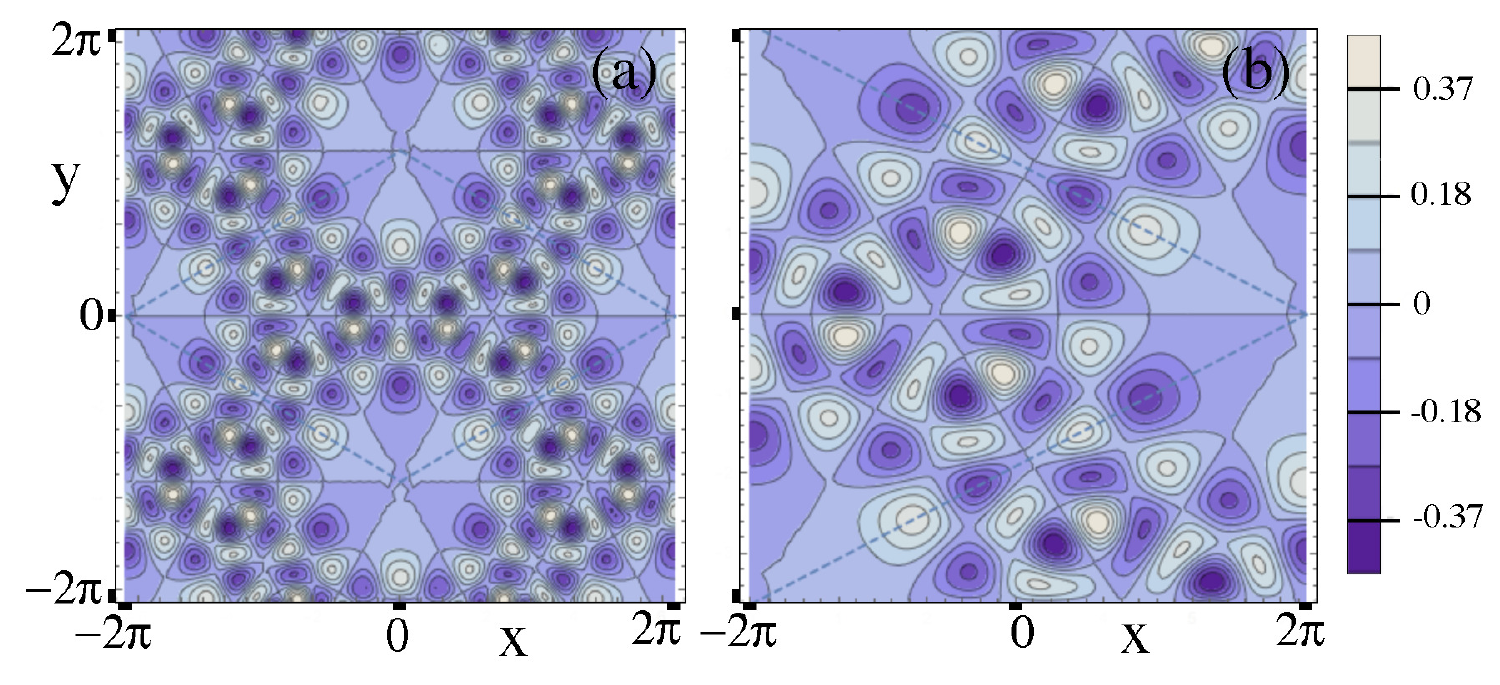}}
\caption{(a) Wave function for an energy eigenstate in the chaotic sea at energy $E=57.8$. The dashed line indicates the unit cell. (b) The same state with focus on the probability amplitude in the half- cell. }
\end{figure}

\end{document}